\documentclass[pra,twocolumn,superscriptaddress,amsmath,amssymb,floatfix,noshowpacs]{revtex4-2} 

\usepackage{graphicx}   
\usepackage{dcolumn}    
\usepackage{bm}         
\usepackage[toc,page]{appendix}

\begin{document}

\title{\textbf{Non-symmetric quantum interfaces with bilayer atomic arrays}}
\author{Roni Ben-Maimon}
\affiliation{Department of Chemical \& Biological Physics, Weizmann Institute of Science, Rehovot 7610001, Israel}
\author{Ofer Firstenberg}
\affiliation{Department of Physics of Complex Systems, Weizmann Institute of Science, Rehovot 7610001, Israel}
\author{Nir Davidson}
\affiliation{Department of Physics of Complex Systems, Weizmann Institute of Science, Rehovot 7610001, Israel}
\author{Ephraim Shahmoon}
\affiliation{Department of Chemical \& Biological Physics, Weizmann Institute of Science, Rehovot 7610001, Israel}
\date{\today}

\begin{abstract}
We study quantum light-matter interfaces based on bilayer atomic arrays in free space, considering interlayer spacings $a_z$ that may deviate from the Bragg-symmetric condition, $a_z\in \mathrm{integer}\times \lambda/2$ with $\lambda$ the light wavelength. Mapping the problem to a one-dimensional model, we show that the interface efficiency is fully determined by simple scattering observables --- reflection and transmission --- providing a direct, experimentally accessible characterization.
This reveals new opportunities for optimizing light-matter coupling by operating beyond the Bragg symmetry. In particular, we identify configurations that suppress diffraction losses via destructive interference, enabling substantially improved interface efficiencies compared to Bragg-constrained designs. In addition, we introduce a new quantum memory scheme based on a collective dark state whose coupling to light is continuously controlled by tuning the interlayer spacing. More broadly, our results establish non-symmetric atomic arrays as a flexible platform for efficient quantum interfaces in free space.
\end{abstract}

\maketitle

\section{Introduction}
Quantum interfaces enable the efficient coupling between atomic excitations and propagating photon modes, underlying a broad range of quantum optical technologies, from quantum memories and networks, to entanglement generation and many-body physics \cite{ref55,ref15,ref17,ref19,ref64}. Recently, it was shown that efficient and directional interfacing to light is established by the collective response of subwavelength atomic arrays \cite{ref28,ref27,Efi2017,ref29,ref30,ref31,ref32,ref37,ref38,ref40,ref18,ref41,ref42,ref43,ref44,ref45,ref46,ref47,ref26,ref48,Robich2}. Of particular importance is the study of such collective effects for overcoming scattering losses in the coupling of light to tweezer atomic arrays \cite{YakovCavity,MultiLayer,MultiLayerMann,Multibeam}, considering the growing prominence of these atomic platforms in quantum research \cite{ref02,ref04,ref05,ref06,ref08,ref09,ref11,ref14,ref65,schlosser2023}.

The efficiency of the quantum interface is generically given by $r_q=\Gamma_q/(\Gamma_q+\gamma_{q,\text{loss}})$, with $\Gamma_q$ being the coupling rates of atomic excitations to a well-defined ``target" optical mode one shines and collects, and $\gamma_{q,\text{loss}}$ describing scattering losses to undesired channels. Fundamentally, $r_q$ is related to the underlying coupling between light and matter as known from classical electrodynamics: this promotes the appealing idea that the efficiency $r_q$ of quantum tasks performed by the interface can be determined from simple scattering observables \cite{Uni}. In disordered ensembles, where backscattering is negligible, this consideration leads to the optical depth as the figure of merit for light–matter coupling \cite{ref51,ref53,ref54}. In contrast, for ordered atomic arrays, where backscattering and multiple scattering play a crucial role, the efficiency was recently shown to be equal to the scattering reflectivity, which can approach unity thanks to the combination of spatial order and cooperative response \cite{Uni,Efi2017}.

The universal characterization of the interface efficiency via the reflectivity applies whenever backward and forward scattering are equal, as can be captured by a symmetric one-dimensional (1D) scattering model of a quantum interface \cite{Uni}. For atomic arrays, such symmetric coupling is realized either in a single 2D array or in a multilayer setup whose interlayer spacing $a_z$ satisfies a Bragg condition, $a_{z}/\lambda\in\mathbb{N}/2$, with $\lambda$ the wavelength of light. In particular, the consideration of multiple array layers offers the advantage of designing destructive (constructive) interference of the lossy modes (target mode) between the different layers. Within the Bragg regime, such an approach was shown to be viable for designing near-unity efficiencies in tweezer arrays by overcoming losses to radiative lattice diffraction orders \cite{MultiLayer,MultiLayerMann}.

\begin{figure}[t]
  \centering
  \includegraphics[width=\columnwidth]{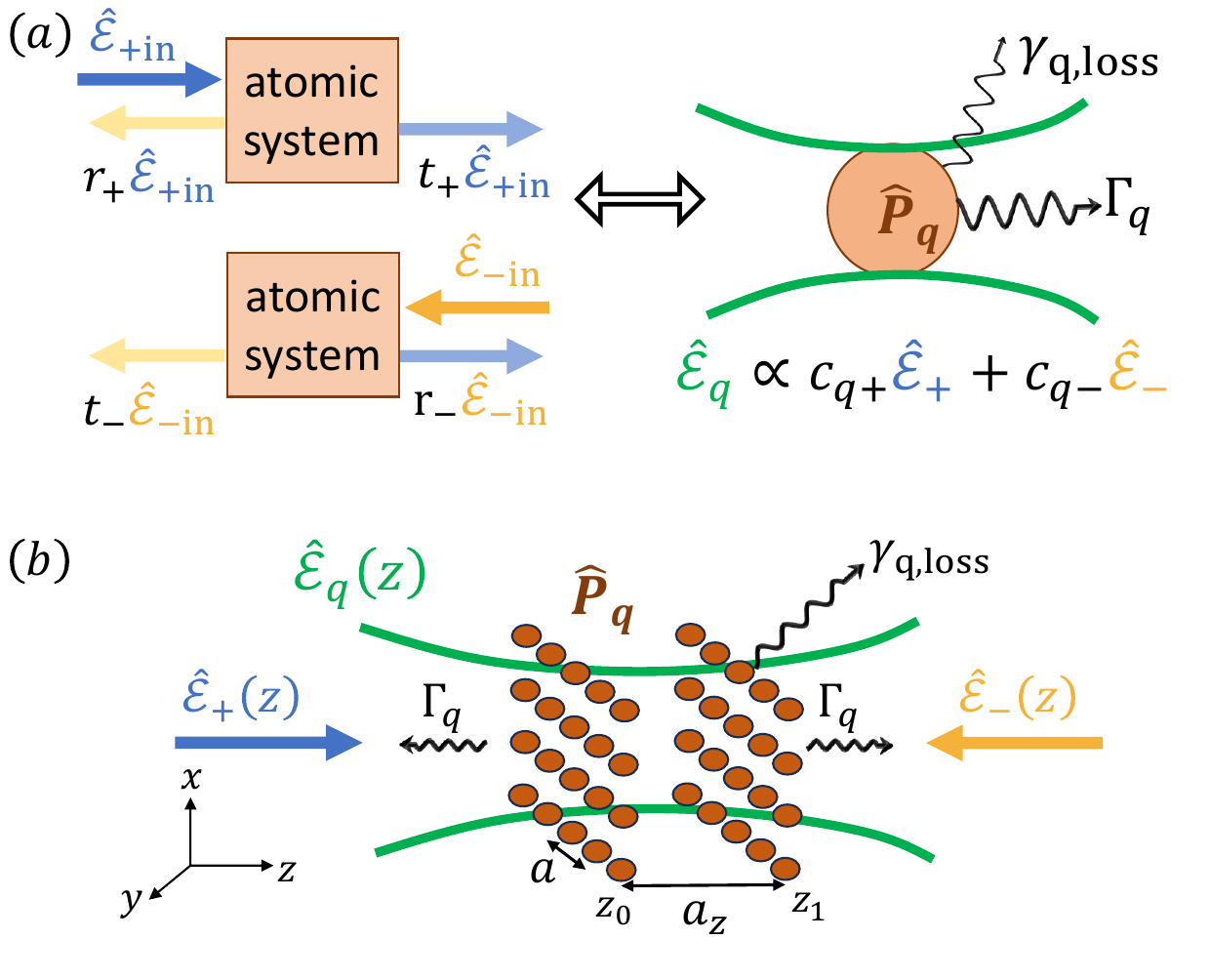}
  \caption{(a) 1D model of a quantum interface. A general scattering problem in 1D is characterized by the reflection and transmission coefficients $r_{\pm}$ and $t_{\pm}$ (left panel). These can be connected to a minimal model of a quantum interface, where the scatterer is a collective atomic excitation $\hat{P}_{q}$ coupled at rate $\Gamma_{q}$ to a spatially matched target photon mode $\hat{\mathcal{E}}_{q}(z)$ (1D propagating), and decays into undesired loss channels at rate $\gamma_{q,\text{loss}}$. The interface efficiency $r_q=\Gamma_{q}/(\Gamma_{q}+\gamma_{q,\text{loss}})$ is fully determined by the scattering observables $r_{\pm}$ and $t_{\pm}$ [Eq. (\ref{rq_ex})]. (b) Realization of the quantum interface using a bilayer atomic array with an arbitrary interlayer spacing $a_z$.}
  \label{Fig1}
\end{figure}

Notably, while the Bragg condition allows for the description of the problem via a single scattering parameter --- the reflectivity --- it also restricts our solutions to be limited to those satisfying only certain interlayer spacings $a_{z}/\lambda\in\mathbb{N}/2$.
This naturally raises the question of the possible advantage of going beyond the Bragg-symmetric case, considering arbitrary lattice spacings and non-symmetric illumination from the opposite sides of the array. Numerical studies of quantum memory in trilayer configurations already suggest that optimizing the phase relation between fields incident from opposite directions can yield efficiencies that surpass those of Bragg-symmetric interfaces \cite{MultiLayerMann}. A systematic study of the advantages of non-symmetric interfaces relevant for a broad class of quantum protocols would then require a universal description of array-based interfaces that explicitly extends beyond Bragg-symmetric configurations.

Here, we develop such a general formulation of a non-symmetric quantum interface composed of a bilayer atomic array with an arbitrary interlayer separation. Whereas Bragg-symmetric interfaces are fully characterized by a single parameter (reflectivity), we show that the non-symmetric regime requires two parameters --- the reflectivity and the transmissivity --- both directly obtainable from classical scattering. We begin by introducing a two-sided 1D scattering model characterized by the reflection and transmission amplitudes $r$ and $t$, respectively, as illustrated in Fig.~\ref{Fig1}a (reducing to the Bragg-symmetric case for $t=1+r$). From this minimal model, we derive the interface efficiency for quantum tasks, $r_{q}$, in terms of the scattering coefficients $r$ and $t$. We then show how the full, many-atom bilayer atomic-array problem [Fig.~\ref{Fig1}b] maps onto this effective 1D model, finding the scattering parameters $r$ and $t$ which determine the efficiency.

We illustrate our formulation for both square and triangular arrays, demonstrating the advantages of non-symmetric interfaces in two applications: (i) enhanced solutions for efficient coupling of tweezer arrays to light, which outperform those that rely on Bragg-symmetric configurations ---
e.g. achieving a five-fold reduction of the inefficiency in considered examples, and enabling the cancelation of higher diffraction-order losses than those attainable in the Bragg regime; (ii) new protocol for quantum memory that does not require three-level atoms, relying on the dynamical control of the coupling of the array to light, $\Gamma\propto 1\pm \cos (2\pi a_z/\lambda)$, via the continuous tuning of the spacing $a_z$ or via a light shift.

\section{1D model of a quantum interface}
\label{secGeneral}
We begin by introducing a minimal model of a two-sided quantum interface to which we later map the atomic array problem. We consider a target photon mode described by 1D field operators propagating either to the right, $\hat{{\cal E}}_{+}(z,t)$ or to the left $\hat{{\cal E}}_{-}(z,t)$, with $[\hat{{\cal E}}_{\pm}(z),\hat{{\cal E}}_{\pm}^{\dagger}(z')]\propto\delta(z-z')$. These fields can be scattered by the atomic system with the corresponding transmission $t_{\pm}$ and reflection $r_{\pm}$ coefficients as (Fig.~\ref{Fig1}a),
\begin{eqnarray}
\hat{{\cal E}}_{\pm}=t_{\pm}\hat{{\cal E}}_{\pm\text{in}}+r_{\mp}\hat{{\cal E}}_{\mp\text{in}},
\label{EIO}
\end{eqnarray}
with $\hat{{\cal E}}_{\pm\text{in}}$ being the corresponding input fields, obeying $[\hat{{\cal E}}_{q,\text{in}}(0,t),\hat{{\cal E}}_{q,\text{in}}^{\dagger}(0,t')]=\delta(t-t')$.

For the analysis of a quantum light-matter interface, we also need to introduce the degrees of freedom of the atomic scatterer, described by the collective-dipole (polarization) operator $\hat{P}_{q}$. This atomic variable can couple to the field from both sides, as captured by the general field superposition,
\begin{eqnarray}
\hat{{\cal E}}_{q}\left(z\right)=\left[c_{q+}\hat{{\cal E}}_{+}\left(z\right)+c_{q-}\hat{{\cal E}}_{-}\left(-z\right)\right]e^{-ikz},
\label{Eq}
\end{eqnarray}
with complex coefficients $c_{q\pm}$ satisfying the normalization $\left|c_{q-}\right|^{2}+\left|c_{q+}\right|^{2}=1$. This light–matter coupling can be modeled by the Heisenberg–picture equations,
\begin{eqnarray}
&&\dot{\hat{P}}_{q}
=\left[-\frac{\Gamma_{q}+\gamma_{q,\text{loss}}}{2}
    + i(\delta-\Delta_{q})\right]\hat{P}_{q}
  + i\sqrt{\Gamma_{q}}\hat{\mathcal E}_{q,\text{in}}(0,t)
+ \hat{F}_{q}(t),\nonumber \\
&&\hat{\mathcal E}_{q}(z)
 =\hat{\mathcal E}_{q,\text{in}}(z)
  + i\sqrt{\Gamma_{q}}\hat{P}_{q}.
\label{EOM}
\end{eqnarray}
Here, $\Gamma_{q}$ is the coupling rate to the target mode $\hat{{\cal E}}_{q}$ while $\gamma_{q,\text{loss}}$ describes coupling to additional loss channels with associated quantum noise $\hat{F}_{q}(t)$. The detuning $\delta$ is between the bare atomic resonance and the central frequency of the target mode, $ck=2\pi c/\lambda$, and $\Delta_{q}$ represents a possible collective dipolar shift. In the low-excitation regime, the collective dipole operator $\hat{P}_{q}$ is linearized and treated as bosonic, $[\hat{P}_{q},\hat{P}_{q}^{\dagger}]=1$, where extensions to the nonlinear regimes are described in \cite{Uni}.

From Eq. (\ref{EOM}), it can be readily shown that the efficiency of conversion between dipolar atomic excitations and target-mode photonic excitations is given by
\begin{eqnarray}
r_{q}=\frac{\Gamma_{q}}{\Gamma_{q}+\gamma_{q,\text{loss}}}.
\label{rq}
\end{eqnarray}
It then follows that the conversion efficiency $r_q$ universally determines the efficiency of a variety of quantum tasks performed by the interface, including quantum memory, entanglement generation and transfer, and more \cite{Uni,ref26,ref48}.

Importantly, $r_q$ also has an analogous meaning in classical physics, as the energy conversion efficiency between dipoles and light. This allows one to express $r_q$ via its relation to the simple scattering observables $r_{\pm}$ and $t_{\pm}$. To show this, we solve classically Eqs. (\ref{EOM}) in steady state considering resonant input field ($\delta=\Delta_q$) coming either from left or right, obtaining [using Eqs. (\ref{EIO}) and (\ref{rq})],
\begin{eqnarray}
r_{q}=\frac{1}{2}\left[1-t_{\pm}-r_{\pm}\left(\frac{c_{q\mp}}{c_{q\pm}}\right)\right].
\label{rq_ex}
\end{eqnarray}
This relation entails the following appealing consequence: once a physical system is mapped onto the quantum interface model of Eq.~(\ref{EOM}), its efficiency in performing quantum-interface operations, $r_q$, is universally determined by a direct measurement or classical calculation of the scattering observables $r_{\pm}$ and $t_{\pm}$ defined in Eq. (\ref{EIO}).

Before we proceed, it is useful to recall the special case of a ``symmetric" interface, wherein the scattering is fully characterized by a single reflectivity parameter. In its simplest form, this corresponds to $r_{+}=r_{-}=r$, $t_{+}=t_{-}=t=1+r$, and $c_{q+}=c_{q-}$. Substituting these relations into Eq.~(\ref{rq_ex}) yields $r_{q}=-r$. Thus, for such a symmetric interface the quantum efficiency is equal to the classical reflectivity, which is the regime analyzed in Ref.~\cite{Uni}. 

\section{Bilayer array quantum interface}
\label{sec_bilayer}
We now turn to the realization of the two-sided quantum interface using a bilayer atomic array, and show how this system maps onto the generalized 1D model introduced in Sec.~\ref{secGeneral}.

\subsection{The system}
We consider the array depicted in Fig. 1b, consisting of two identical layers with an interlayer spacing $a_{z}$ along the $z$ axis. Each layer forms a lattice in the $xy$-plane characterized by the lattice spacing $a$ and an angle $\psi$ between the primitive lattice vectors, taking the values $\psi=\frac{\pi}{2}$ and $\frac{\pi}{3}$ for square and triangular lattices, respectively. The corresponding atomic positions are given by $\mathbf{r}_{\mathbf{n},l}=(n_1 a+n_2a\cos\psi,n_2 a\sin\psi ,z_l)$, where $\mathbf{n}=(n_{1},n_{2})$ labels the lattice sites within each layer and $l=1,2$ denotes the layer index, with $z_l-z_{l'}=(l-l')a_z$. The arrays are illuminated by a normal incident beam of central wavelength $\lambda=2\pi/k$; for $a<\lambda$ the system operates in the subwavelength regime, whereas for $a>\lambda$ it lies in the superwavelength regime.

Beginning with the full problem of two-level atoms (ground and excited levels $|g\rangle$ and $|e\rangle$) coupled to free-space photon modes, we eliminate the photon modes within a Born-Markov approximation, obtaining the many-atom Heisenberg–Langevin equations for the lowering atomic operators $\left|g\right\rangle \left\langle e\right|_{\mathbf{n}l}$ coupled to each other via photon-mediated dipole–dipole interactions under weak illumination (linear response) \cite{Uni}. For simplicity, we consider first that each layer forms an infinite 2D lattice, and account for finite-size effects further below. In this case, the many-atom equations can be simplified by defining the symmetric collective dipole operator of a layer $l$, $\hat{P}_{l}\propto \sum_{\mathbf{n}}\left|g\right\rangle \left\langle e\right|_{\mathbf{n}l}$ that is naturally coupled to the normal-incident field. This results in the following coupled equations between the layers,
\begin{eqnarray}
&&\dot{\hat{P}}_{l}
  = \left(i\delta-\gamma_{s}\right)\hat{P}_{l}
   - \sum_{l'=1}^{2}D_{ll'}\hat{P}_{l'}
\nonumber\\
   &&+ i\sqrt{\frac{\Gamma_{\text{1D}}}{2}}
     \left[\hat{\mathcal E}_{+\text{in}}(z_{l})
     + \hat{\mathcal E}_{-\text{in}}(z_{l})\right]
   + \hat{F}_{l},
\nonumber\\
&&\hat{\mathcal E}_{\pm}(z)
  = \hat{\mathcal E}_{\text{in}\pm}(z)
   + i\sqrt{\frac{\Gamma_{\text{1D}}}{2}}\,e^{\pm ikz}
     \sum_{l=1}^{2} e^{\mp ikz_{l}} \hat{P}_{l}.
\label{EOM_layer}
\end{eqnarray}
Here, $\Gamma_{\text{1D}}=\frac{3\gamma}{4\pi\sin\psi}(\frac{\lambda}{a})^{2}$ denotes the collective emission rate of a single layer into the normal-incident target mode $\hat{\mathcal E}_{\pm}(z)$, with $\gamma$ being the single-atom spontaneous-emission rate in free space.
The decay rate $\gamma_s$ accounts for the possibility of additional loss processes at the individual-atom or individual-layer level that go beyond the perfect infinite-layer two-level atom arrays taken here, such as due to additional ground states, weak disorder or finite-size effects (see below). Accordingly, $\hat{F}_{l}$ is the quantum Langevin noise operator associated with these decay processes and with the collective emission to lattice diffraction orders beyond the normal-incident target mode (see below).

The layers are coupled by the effective dipole–dipole interaction kernel, obtained from the projection of the photon Green's function onto the layer dipole modes $\hat{P}_l$,
\begin{eqnarray}
D_{ll'}=\sum_{\mathbf{m}}\frac{\Gamma_{\mathbf{m}}}{2}e^{ik_{z}^{\mathbf{m}}a_{z}\left|l-l'\right|}.
\label{D_ll}
\end{eqnarray}
This includes contributions from all of the diffraction orders $\mathbf{m}=(m_{1},m_{2})$ of the 2D lattice that forms each layer, with $m_{1},m_{2}\in\mathbb{Z}$ and corresponding transverse wavevectors (reciprocal lattice vectors) $\boldsymbol{Q}_{\mathbf{m}}=\frac{2\pi}{a}(m_{1},-m_{2}\cot\psi+m_{2}/\sin\psi)$. The coupling rate of the layer collective dipole $\hat{P}_l$ to a diffraction order $\mathbf{m}$ is
\begin{eqnarray}
\Gamma_{\mathbf{m}}=\Gamma_{\text{1D}}\frac{1-\left|\boldsymbol{Q}_{\mathbf{m}}\cdot\mathbf{e}_{\mu}\right|^{2}/k^{2}}{k_{z}^{\mathbf{m}}/k},
\quad
k_{z}^{\mathbf{m}}=\sqrt{k^2-\left|\boldsymbol{Q}_{\mathbf{m}}\right|^{2}},
\label{Gm}
\end{eqnarray}
where $\mathbf{e}_{\mu}$ denotes the atomic dipole orientation (taken to be circularly polarized). The zeroth order $\mathbf{m}=\boldsymbol{0}$ reproduces the usual interaction kernel of QED in 1D, $\Gamma_{\text{1D}}e^{ika_{z}\left|l-l'\right|}/2$. The quasi-1D contributions of the higher diffraction orders $\left|\mathbf{m}\right|\neq\boldsymbol{0}$ can be divided into evanescent ($k_{z}^{\mathbf{m}}\in \text{Im}$) and radiative ($k_{z}^{\mathbf{m}}\in\text{Re}$), the latter satisfying
\begin{eqnarray}
\frac{\left|\boldsymbol{Q}_{\mathbf{m}}\right|}{k}=\left|\frac{\lambda}{a}(m_{1},-m_{1}\cot\psi+\frac{m_{2}}{\sin\psi})\right|<1,
\label{R}
\end{eqnarray}
and denoted by the set $\mathbf{m}\in R$. In the subwavelength regime, only the zeroth diffraction order $\mathbf{m}=(0,0)$ is radiative. In contrast, in the superwavelength regime, and as the lattice spacing $a$ gets larger, an increasing number of radiative diffraction orders can satisfy the condition (\ref{R}) and become radiative, leading to additional scattering losses beyond the individual decay $\gamma_s$.

\subsection{Mapping to the 1D model}
We can simplify Eqs. (\ref{EOM_layer}) by writing them using the collective-dipole eigenmodes of the bilayer interaction kernel $D_{ll'}$ from Eq. (\ref{D_ll}),
\begin{eqnarray}
\hat{P}_{q}=\frac{1}{\sqrt{2}}\left[\hat{P}_{1}+e^{iq}\hat{P}_{2}\right],
\label{Pq}
\end{eqnarray}
with $q=0,\pi$ corresponding to the in-phase and out-of-phase superposition of the layers, respectively. The collective dipoles $q=0,\pi$ are then seen to couple to corresponding photonic fields in the form of Eq. (\ref{Eq}), with coefficients
\begin{eqnarray}
c_{0,\pm}=\frac{e^{\pm ikz_{c}}}{\sqrt{2}},\quad c_{\pi,\pm}=\mp i\frac{e^{\pm ikz_{c}}}{\sqrt{2}},
\label{cq}
\end{eqnarray}
with $z_{c}=(z_{1}+z_{2})/2$ denoting the midpoint between the two layers. Notably, the corresponding photonic target modes $q$ describe fields arriving at the bilayer midpoint from both sides either with the same ($q=0$) or with opposite ($q=\pi$) phases.

In terms of the above bilayer collective dipoles $\hat{P}_q$ and corresponding target modes $\hat{{\cal E}}_{q}$, Eqs. (\ref{EOM_layer}) take the form of the 1D model, Eq. (\ref{EOM}), with the model parameters
\begin{align}
\Gamma_{q}=\Gamma_{\text{1D}}\left[1+e^{iq}\cos\left(ka_{z}\right)\right],\quad\gamma_{q,\text{loss}}=\gamma_{q,\text{diff}}+\gamma_{s}.
\label{effParamarray}
\end{align}
Here the loss term $\gamma_{q,\text{diff}}$ includes all contributions from coupling to higher diffraction orders that are radiative,
\begin{eqnarray}
\gamma_{q,\text{diff}}=\sum_{\substack{\mathbf{m}\in R \\ \mathbf{m}\neq\mathbf{0}}}\Gamma_{\mathbf{m}}\left[1+e^{iq}\cos\left(k_{z}^{\mathbf{m}}a_{z}\right)\right].
\label{G_diff}
\end{eqnarray}
A collective energy shift $\Delta_{q}$ is also obtained, with contributions from both radiative and non-radiative orders,
\begin{align}
\Delta_{q}
  = e^{iq} \sum_{\mathbf{m}\in R} \frac{\Gamma_{\mathbf{m}}}{2}
    \sin\!\left(k_{z}^{\mathbf{m}}a_{z}\right)
  + \sum_{\mathbf{m}\notin R} \frac{|\Gamma_{\mathbf{m}}|}{2}
    \left[1 + e^{iq} e^{-|k_{z}^{\mathbf{m}}|a_{z}}\right].
\label{Delta_q}
\end{align}

After establishing the mapping to the 1D model, the interface efficiency $r_q$ of each collective mode $q=0,\pi$ can be fully determined, either from theory using the model parameters from Eq. (\ref{effParamarray}) in Eq. (\ref{rq}), or directly from the scattering observables $t_{\pm}$ and $r_{\pm}$ via Eq. (\ref{rq_ex}). For the latter, we find in our bilayer case that the required scattering observables reduce to only two quantities $t=t_{+}$ and $r=r_{+}$, by noting the relation [readily derived from Eqs. (\ref{rq_ex}) and (\ref{cq})],
\begin{align}
t=t_{+}=t_{-},\quad r=r_{+}=r_{-}e^{4ikz_{c}},
\end{align}
recalling that $z_c$ denotes the midpoint between the layers and effectively serves as a ``gauge" choice. If one further considers the Bragg-symmetric case, then one obtains the
additional relation $t=1+re^{ik(a_{z}-2z_{c})}$, such that the efficiency is determined by the single reflectivity parameter $r$, as $r_q=|r|$, which is the essence of the symmetric interface \cite{Uni}. Choosing the gauge $z_{c}=a_{z}/2$, this takes the familiar form $t=1+r$ known for single-layer scatterers. In this symmetric-interface case, one collective mode is fully bright and the other is fully dark, with the bright mode being coupled solely to in-phase illumination.

\subsection{Coupling rate and efficiency}
The coupling efficiency $r_{q}$ is basically determined by the interplay between the target mode coupling $\Gamma_{q}$ and loss $\gamma_{q,\text{loss}}$. Starting with the former, we note that the coupling rate $\Gamma_{q}$ is controlled by the interlayer spacing $a_{z}$ [Eq. (\ref{effParamarray})], such that the collective modes $q=0,\pi$ alternate between strongly and weakly coupled (super- and sub-radiant), turning into completely bright or dark modes within the Bragg condition. We show in Sec. \ref{sec_qmem} below how tuning between dark and bright conditions by varying $a_z$ can serve as the basis for a quantum memory scheme.
Meanwhile, the losses $\gamma_{q,\text{loss}}$ strongly depend on the intralayer lattice spacing $a$. For subwavelength arrays ($a<\lambda$), we have $\gamma_{q,\text{loss}}=\gamma_{s}$, such that the loss rate arises solely from non-collective processes, which are typically small compared with $\Gamma_{\text{1D}}$, leading to high coupling efficiency \cite{Efi2017,Uni}. Conversely, in superwavelength arrays ($a>\lambda$), the emergence of additional radiative diffraction orders leads to scattering losses $\gamma_{q,\text{diff}}$ that can overwhelm $\Gamma_{\text{1D}}$, resulting in much reduced coupling efficiency.

\section{Application 1: Tweezer array quantum interface}
As discussed in the introduction, the increasing prominence of tweezer atomic arrays in quantum science emphasizes the need and possible benefits of their efficient interface with light. This endeavor is however severely hampered due to the typical operation of tweezer arrays in the superwavelength regime ($a>\lambda$), where the losses due to scattering to higher diffraction orders, $\gamma_{q,\text{diff}}$, become substantial.

Recent works have shown how the bilayer scheme can be harnessed to reduce these losses by designing their interlayer destructive interference \cite{MultiLayer,MultiLayerMann}. More specifically, by establishing the mapping to the 1D model in the Bragg regime $a_{z}/\lambda\in\mathbb{N}/2$, a formula for $\gamma_{q,\text{diff}}$ in the form of Eq. (\ref{G_diff}) was derived and conditions for the cancellation between its two terms were obtained: these conditions yield discrete ``resonant sets” of spacings $(a_{z},a)$ for which $\gamma_{q,\text{diff}}= 0$, resulting in near-unity interface efficiencies. While this represents a viable approach, it is inherently limited: the Bragg condition restricts $a_{z}$ to specific values $\mathbb{N}
\lambda /2$ that lead to corresponding restricted values of $a$. This lack of flexibility in choosing $(a_{z},a)$ limits the achievable efficiency in realistic finite-size arrays, as discussed further below.

Notably, however, the restriction to these discrete values is not fundamental and originates merely in the consideration of Bragg-symmetric interfaces. In contrast, our general non-symmetric formulation allows to extend the basic idea of destructively interfering diffraction beyond the restrictive Bragg case. In particular, by again demanding $\gamma_{q,\text{diff}}=0$ in Eq. (\ref{G_diff}), we now reveal a \emph{continuous} family of ``resonant curves" $(a_{z},a)$ of high efficiency: this allows to easily identify configurations that substantially outperform the restricted Bragg-symmetric solutions in realistic finite-size arrays.

As the intralayer spacing $a$ increases, additional diffraction orders become radiative and must be cancelled to maintain high efficiency. Below, we demonstrate how such cancellations can be achieved, beginning with the first and then the second radiative diffraction orders, for both square and triangular array geometries.

\subsection{Eliminating diffraction losses on resonant curves}
\begin{figure}[t]
  \centering
  \includegraphics[width=\columnwidth]{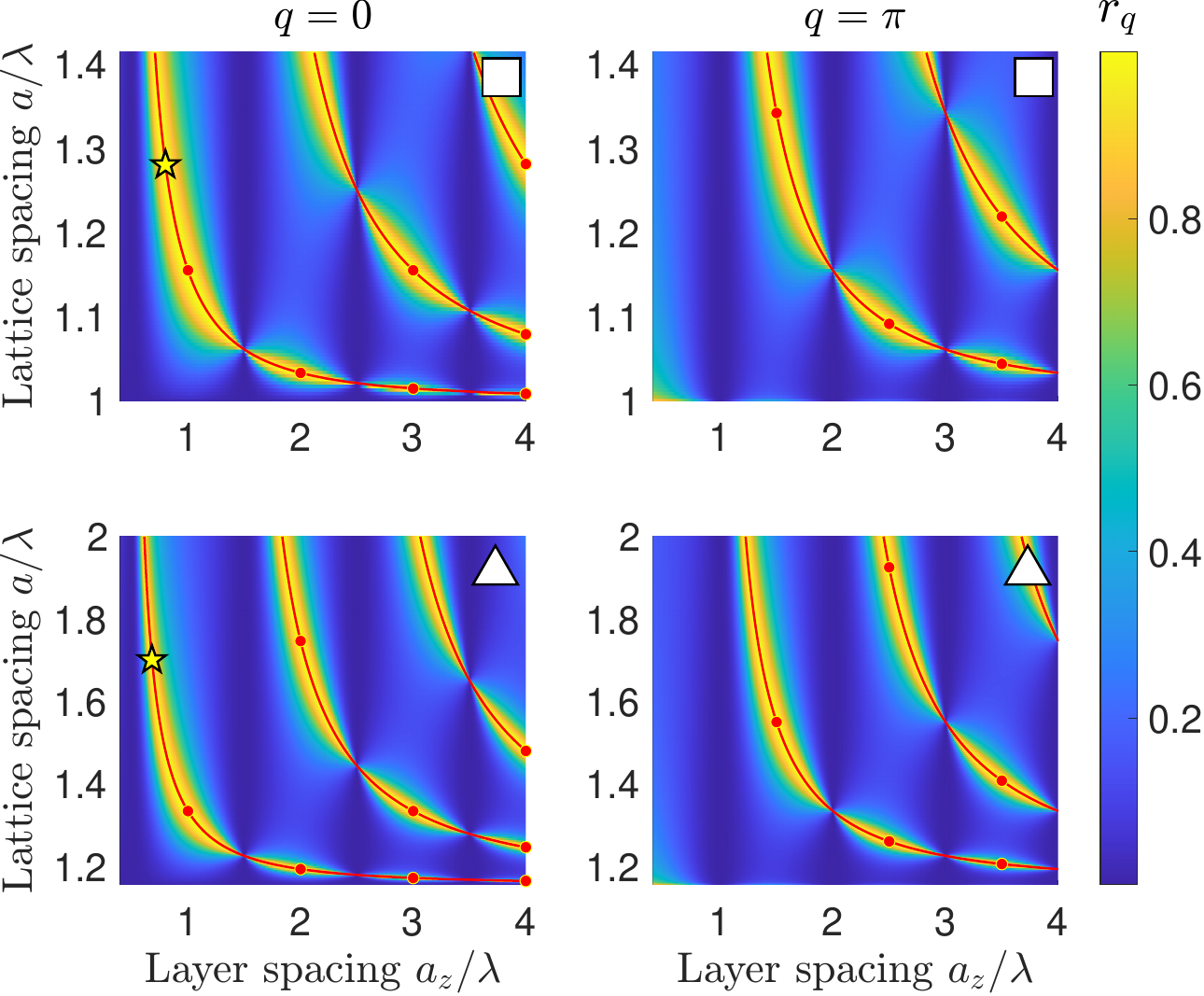}
  \caption{Efficiency $r_{q}$ of the in-phase ($q=0$) and out-of-phase ($q=\pi$) modes of the bilayer-array quantum interface, as a function of the interlayer spacing $a_z$ and intralayer spacing $a$. Results are obtained analytically from Eq. (\ref{rq}) with the effective parameters from Eqs. (\ref{effParamarray}) and (\ref{G_diff}) (with $\gamma_s=0$), considering infinite superwavelength ($a>\lambda$) layers, forming either square (\(\square\)) or triangular (\(\triangle\)) lattices. Regions of optimal efficiency $r_{q}= 1$ are seen to coincide with the ``resonant curves" from Eq. (\ref{curves}), marked in red lines. Red points indicate the set of optimal configurations $(a_z,a)$ along the resonant curves, which satisfy the Bragg condition, while the stars indicate the configurations analyzed in  Fig.~\ref{Fig3}.}\label{Fig2}
\end{figure}

The radiative diffraction losses can be eliminated by appropriately choosing the array spacings $(a_{z},a)$ such that the contributions of all radiative diffraction modes $\mathbf{m}\in R,\neq\mathbf{0}$ in Eq. (\ref{G_diff}) interfere destructively, satisfying $e^{iq}\cos(k_{z}^{\mathbf{m}}a_{z})=-1$. We begin by examining the superwavelength regime in which only the first diffraction order is radiative, corresponding to $a/\lambda\in[1,\sqrt{2}]$ for square and $a/\lambda\in[2/\sqrt{3},2]$ for triangular lattices (wherein diffraction modes $\mathbf{m}=\left\{ (\pm1,0),(0,\pm1)\right\}$ for square and $\mathbf{m}=\left\{ (\pm1,0),(0,\pm1),(\pm1,\pm1)\right\}$ for triangular become radiative). In this regime, solving the interference condition yields continuous resonant curves in the $(a,a_{z})$ plane for which $\gamma_{q,\text{diff}}=0$,
\begin{eqnarray}
\frac{a}{\lambda}=
\left\{
  \begin{array}{ll}
    \left[1-(\lambda/a_z)^2\left(n_c+\frac{1-q}{2}\right)^{2}\right]^{-1/2}, & \hbox{\text{square};} \\
    \frac{2}{\sqrt{3}}\left[1-(\lambda/a_z)^2\left(n_c+\frac{1-q}{2}\right)^{2}\right]^{-1/2}, & \hbox{\text{triangular},}
  \end{array}
\right.
\nonumber\\
\label{curves}
\end{eqnarray}
with $n_c\in\mathbb{N}$ labeling the different resonant curves. Figure \ref{Fig2} presents the theoretical map of the coupling efficiency $r_{q}$ as a function of $(a_{z},a)$, obtained from Eqs. (\ref{rq}), (\ref{effParamarray}) and (\ref{G_diff}) for $\gamma_s=0$. It reveals that high efficiencies are indeed exhibited on the continuous resonant curves, Eq. (\ref{curves}), for both lattice geometries and under both in- and out-of-phase illumination $q=0,\pi$ (noting the subtlety of the Bragg solutions, c.f. \cite{note}).

\begin{figure}[t]
  \centering
  \includegraphics[width=\columnwidth]{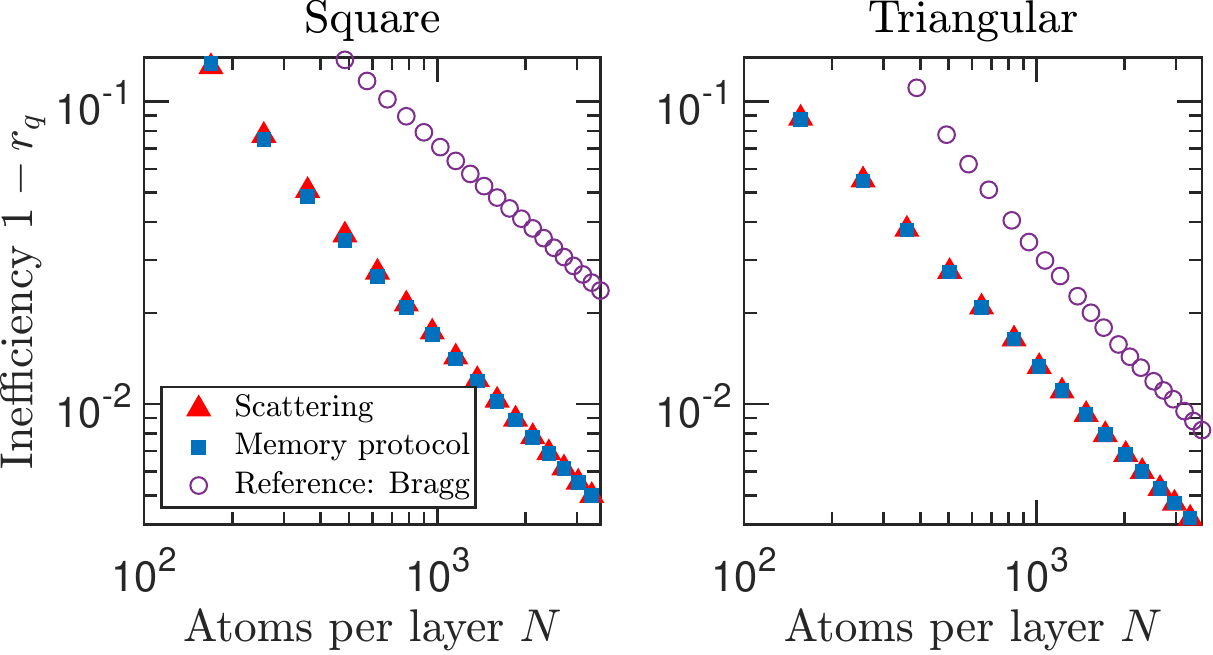}
  \caption{Finite-size scaling in the bilayer-array quantum interface: coupling inefficiency $1-r_{q}$ as a function of the atom number per layer $N$. Results of direct numerical scattering simulations of a normal-incident Gaussian-beam with waist $w/L=0.26$ ($L\sim a \sqrt{L}$) are presented in red triangles; $r_{q}$ is evaluated from the calculated reflectivity and transmissivity using Eq.~(\ref{rq_ex}), for square and triangular arrays with spacings $\ensuremath{(a_{z},a)}/\lambda=(0.80,1.28)$ and $\ensuremath{(a_{z},a)}/\lambda=(0.68,1.70)$, respectively (marked by stars in Fig. 2). The resulting inefficiency $1-r_{q}$ scales favorably as $\sim N^{-1}$ (fits for $N>360$ yield $N^{-1.03}$ and $N^{-0.99}$ for square and triangular, respectively). Excellent agreement is observed with direct calculations of the infidelity of a quantum memory protocol performed for the same Gaussian-beam target mode (blue squares), confirming that $r_{q}$ quantifies efficiency of quantum tasks. The reference case of the Bragg-symmetric arrays, with the optimal configurations studied in Ref. \cite{MultiLayer} [$\ensuremath{(a_{z},a)}/\lambda=(3.0,1.342)$ and $\ensuremath{(a_{z},a)}/\lambda=(2.5,1.925)$ arrays] shows similar scalings with $N$ but worse overall performance (circles).}
  \label{Fig3}
\end{figure}

So far we have considered infinite atomic layers; we now show how these ideas apply to realistic finite-size atomic layers and target-mode beams. To this end, we perform classical wave scattering simulations considering a finite-waist Gaussian-beam target mode at resonance ($\delta=\Delta_q$), incident on the finite-size array \cite{Efi2017}. Calculating the forward- and backward-scattered fields and projecting them onto the Gaussian target-mode profile, we extract the scattering amplitudes $t$ and $r$, respectively, from which the coupling efficiency $r_{q}$ is obtained via Eq. (\ref{rq_ex}).
All simulations are performed with a fixed beam waist $w/L=0.26$ ($L\sim\sqrt{N}a$ being the layer size), which serves as a near-optimal compromise between wave-diffraction losses and spatial beam–layer mismatch (dominant for small and large beam waists, respectively \cite{YakovCavity}). Figures \ref{Fig3}a-b present the resulting coupling inefficiency $1-r_{q}$ as a function of the atom number per layer $N$, for the resonant-curve configurations $\ensuremath{(a_{z},a)}/\lambda=(0.80,1.28)$ and $\ensuremath{(a_{z},a)}/\lambda=(0.68,1.70)$ of the in-phase mode $q=0$ of square and triangular arrays, respectively (noting that other configurations exhibit similar results). The simulations reveal a universal scaling across different configurations, $1-r_{q}\propto N^{-1}$. Considering the general form of $r_q$ from Eq. (\ref{rq}), this result is interpreted as a residual individual-like loss channel, $\gamma_{q,\text{loss}}=\gamma_s \propto N^{-1}$, originated in imperfect cancellation of diffraction-order losses between finite-size layers.

\begin{figure}[t]
  \centering
  \includegraphics[width=\columnwidth]{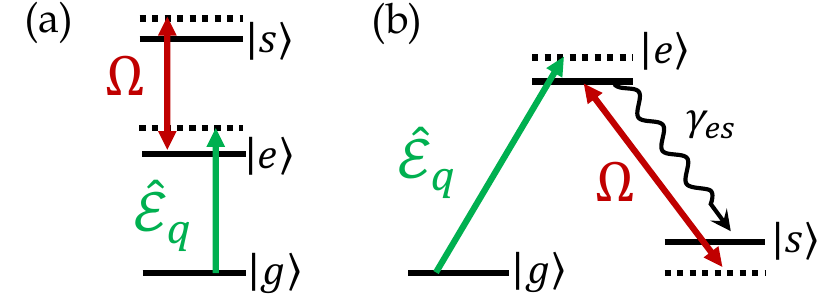}
  \caption{Three-level atom quantum memory: ladder-type (a) and $\Lambda$-type (b) schemes. The photon field $\hat{{\cal E}}_{q}$ excites the collective dipole $\hat{{P}}_{q}$, which is coupled via the control field $\Omega$ to the collective coherence between the long-lived states $\left|g\right\rangle$ and $\left|s\right\rangle$, where excitations are stored. In the $\Lambda$ scheme an additional individual decay rate $\gamma_{es}$ from $|e\rangle$ to $|s\rangle$ is added to the loss $\gamma_{q,\text{loss}}$ [via an individual loss term $\gamma_s=\gamma_{es}$ in Eq. (\ref{effParamarray})].
}\label{Fig6}
\end{figure}

In addition to the definition of $r_q$ via scattering observables, we directly demonstrate its operational meaning as the efficiency of quantum tasks performed by the array interface, as predicted by the mapping to the 1D model. To this end, we take the example of a typical quantum memory protocol where photonic excitations are mapped onto collective excitations of the coherence between two stable atomic states $|g\rangle$ and $|s\rangle$, either in a ladder- or a $\Lambda$-type three-level-atom configuration (Fig.~\ref{Fig6}) \cite{gorshkovCavMem,ref51,FL}. Using the method from Ref.~\cite{ref29}, we perform direct numerical simulations of this memory protocol for the bilayer array of such three-level atoms.  The results for the memory infidelity using the same array geometries and target-mode illumination as in the scattering simulations are presented in Fig. \ref{Fig3} (blue squares), exhibiting excellent agreement with the scattering results for $1-r_{q}$.

Notably, our calculations of the memory protocol assume a ladder configuration (Fig.~\ref{Fig6}a), recalling that in practice the additional state $|s\rangle$ is an excited quasi-stable state whose finite lifetime limits the memory storage time (e.g. a Rydberg state with a lifetime of $10-100$ $\mu$s). In contrast, the $\Lambda$ configuration offers a stable ground state $|s\rangle$, however, at the cost of a loss term $\gamma_s=\gamma_{es}$ added to $\gamma_{q,\text{loss}}$ in Eq. (\ref{effParamarray}), which is originated in the individual-atom decay $\gamma_{es}$ from $|e\rangle$ to $|s\rangle$ (Fig.~\ref{Fig6}b). This would add a constant $\gamma_s/(\Gamma_q+\gamma_{q,\text{loss}})\sim \gamma_{es}/\Gamma_{\text{1D}}$ to the inefficiency $1-r_q$ plotted in Fig. 3 [Eq. (\ref{rq})]. In Sec.~\ref{sec_qmem} below, we introduce an alternative quantum memory scheme based on two-level atoms, avoiding these limitations associated with three-level atoms.

\subsection{Enhanced efficiencies in non-symmetric interfaces}

While all the configurations $(a_z,a)$ on the resonant curves guarantee equally perfect efficiency $r_q=1$ for infinite arrays, their different performance in realistic finite-size cases, seen in Fig. 3, can be understood from a geometrical-optics picture \cite{MultiLayer}. In particular, light scattered into the $\mathbf{m}\neq 0$ diffraction orders propagates at angles $\theta_{d}=\arcsin(\left|\boldsymbol{Q}_{\mathbf{m}}\right|/k)$ and, in finite arrays, undergoes only a limited number of reflections between the two layers before escaping the bilayer structure. These escaping rays do not contribute to the overall, normal-incident target-mode field and hence constitute the dominant loss mechanism in realistic systems. Within this picture, the inefficiency decreases with increasing array size, as captured by the result $1-r_q \sim M^{-1}$ with $M\sim(\sqrt{N}a/a_{z}\tan\theta_{d})^{2}$ \cite{MultiLayer}. This explains the scaling $N^{-1}$ seen in the numerical results of Fig. 3.

The geometrical-optics picture provides a practical design principle: minimizing ray-escape losses favors smaller interlayer spacings $a_{z}$ and smaller diffraction angles $\theta_{d}$, the latter corresponding to larger intralayer spacings $a$. Along any resonant curve, the optimal operating point therefore lies at the largest possible $a$ values, near the onset of the next radiative order. Nevertheless, choosing $a$ too close to this edge becomes counterproductive: finite beam waist introduces momenta components that leak into the next radiative order and increase losses \cite{Multibeam}. The configurations chosen in Fig.~\ref{Fig3} optimize this trade-off for the range of $N$ considered at the chosen (fixed) beam waist.

Importantly, considering non-symmetric interfaces beyond the Bragg condition, allows for the optimization of the efficiency $r_q$, as per the above design principle, over a continuous locus of solutions $(a_z,a)$ along resonant curves. This is in contrast to the Bragg-symmetric case, which offers only a discrete subset of configurations to choose from, out of the continuous resonant curve solutions (Fig. 2, red dots). Therefore, the Bragg-restricted solutions are rarely the optimal ones: in Fig.~3 we plot the best Bragg-restricted solutions (circles), which are seen to perform significantly worse than the optimal solutions found by lifting the Bragg restrictions --- with up to a factor 5 difference in the simulated configurations. This directly demonstrates the benefits of considering array geometries beyond the Bragg-symmetric case, using the approach developed here.

\subsection{Eliminating higher diffraction orders}
Extending the analysis to regimes where two diffraction orders are radiative enables operation at larger lattice spacings beyond the first-order limit. This requires simultaneous cancellation of the contributions of both the first and the second radiative orders, to maintain high efficiency. Within a Bragg-constrained bilayer array, such exact cancellation is impossible, since for any Bragg-fixed spacing $a_{z}$ the two phase matching conditions cannot be satisfied simultaneously, allowing only approximate cancellation \cite{MultiLayer}.

\newcommand{\squaredot}{%
  \mathbin{\ooalign{$\square$\cr\hidewidth\raisebox{0.25ex}{$\bullet$}\hidewidth\cr}}%
}

\begin{figure}[t]
  \centering
  \includegraphics[width=\columnwidth]{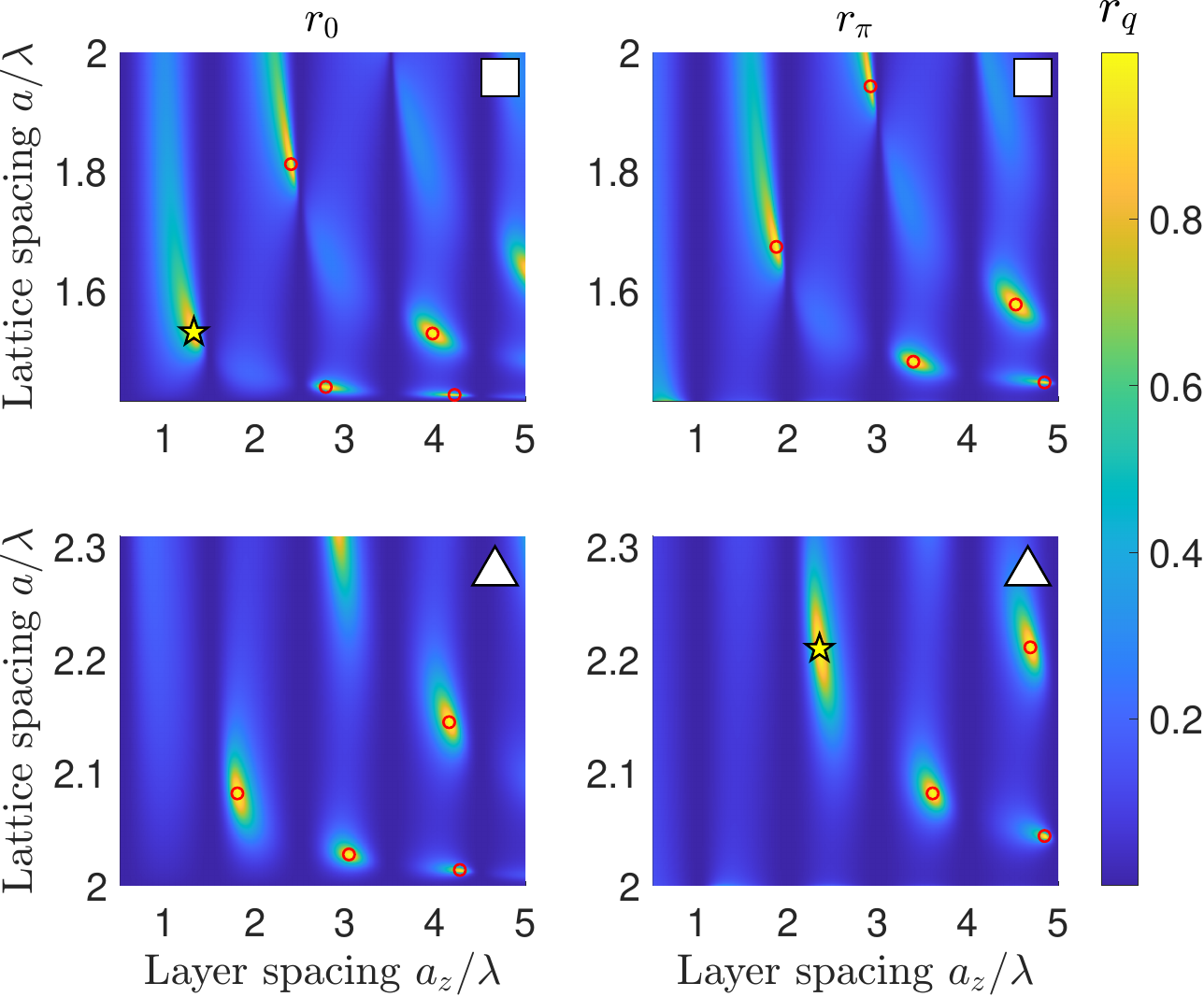}
  \caption{Eliminating higher diffraction orders: resonant sets. Efficiency $r_{q}$ as a function of the interlayer spacing $a_z$ and intralayer lattice spacing $a$ obtained analytically from Eqs. (\ref{rq}), (\ref{effParamarray}) and (\ref{G_diff}), considering the regime where two diffraction orders are radiative (infinite layers, $\gamma_s=0$). The square array configurations (\(\square\)) include a lateral shift $(a/2,a/2)$ between the layers. Resonant sets $(a,a_{z})$ of high-efficiency predicted from the condition $\gamma_{q,\text{diff}}=0$ are marked in red circles and coincide with $r_{q}= 1$ as expected. Stars indicate the configurations studied in Fig.~\ref{Fig5}.}\label{Fig4}.
\end{figure}

By contrast, removing the Bragg constraint relaxes the geometric requirements and enables exact two-order cancellation using only two layers. Solving $\gamma_{q,\text{diff}}=0$ imposes simultaneous phase-matching conditions on both radiative orders and yields discrete resonant sets $(a_{z},a)$ of exact two-order cancellation and resulting high interface efficiency. These resonant sets are marked on the efficiency map plotted as a function of $(a_z,a) $ in Fig. \ref{Fig4} for the regime of two radiative orders; namely $a/\lambda\in[\sqrt{2},2]$ for the square and $a/\lambda\in[2,4/\sqrt{3}]$ for the triangular lattice. Indeed, the predicted resonant sets are seen to coincide with points of high efficiency $r_q$. Notably, for the square array configurations, a lateral shift of $(a/2,a/2)$ was introduced between the two layers, providing an additional interference pathway that generates new resonant solutions. .

\begin{figure}[t]
  \centering
  \includegraphics[width=\columnwidth]{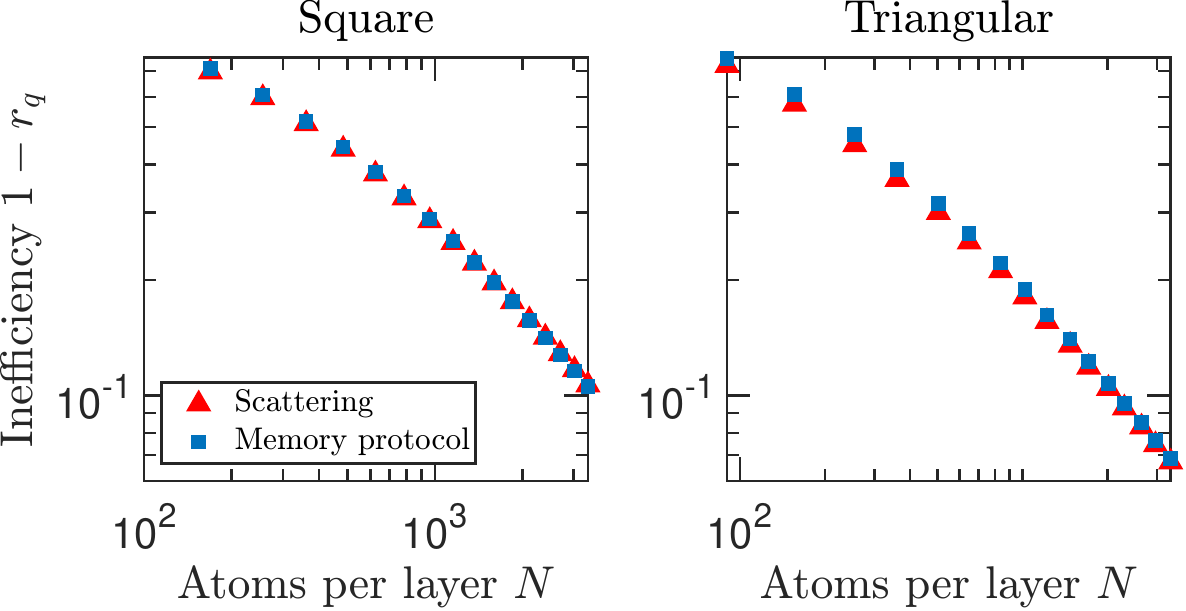}
  \caption{Eliminating higher diffraction orders: finite-size scaling.
  Coupling inefficiency $1-r_{q}$ as a function of the atom number per layer $N$, evaluated using Eq.~(\ref{rq_ex}) from the reflection and transmission obtained via direct scattering numerical simulations (beam waist $w/L=0.26$). Results for square and triangular arrays (red triangles) with the optimal spacings $\ensuremath{(a_{z},a)}/\lambda=(1.53,1.32)$ and $\ensuremath{(a_{z},a)}/\lambda=(2.35,2.21)$, respectively (marked by stars in Fig.~\ref{Fig4}), exhibit a decrease with fitted scalings $N^{-0.80}$ and $N^{-0.83}$ (for $N\geq10^{3}$), approaching the favorable asymptotic scaling $1-r_{q}\sim N^{-1}$. Results of infidelity of the quantum memory protocol exhibits excellent agreement (blue squares).}\label{Fig5}
\end{figure}

To assess the performance of realistic interfaces, we evaluate the coupling inefficiency $1-r_{q}$ as a function of the atom number per layer $N$ for the optimal configurations identified in Fig. \ref{Fig4}. The results, displayed in Fig. \ref{Fig5}, show that $1-r_{q}$ again decreases favorably with $N$, exhibiting a slightly weaker power-law than the asymptotic $N^{-1}$ scaling, within the simulated range of $N$ values. This is attributed to the larger diffraction angles $\theta_{d}$ associated with second-order radiative modes, leading to an increased escape of rays.

\section{Application 2: Quantum memory with two level atoms}
\label{sec_qmem}
Quantum memories are based on the ability to coherently transfer quantum excitations between light and stable atomic degrees of freedom, by the dynamical control of their coupling to light. Conventionally, the latter relies on finding suitable coherent and metastable electronic states of \emph{individual} atoms, as in the three-level atom schemes from Fig.~\ref{Fig6}. While the overall quantum memory is formed by an ensemble of such atoms, the limitations on memory performance of these schemes originate in individual-atom physics; specifically, in the lifetime and coherence associated with the auxiliary single-atom metastable state $|s\rangle$, considering the ladder and $\Lambda$ schemes, respectively (recalling Sec.  IV A).

In contrast, we now present a quantum-memory approach that relies on the emergence of metastable states at the \emph{collective} level, originating in subradiant collective excitations of atomic arrays.  Here, only the two levels $|g\rangle$ and $|e\rangle$ of each atom are used, without the need for auxiliary single-atom metastable states.
Such a memory scheme was introduced recently for a single-layer
subwavelength 2D array \cite{YelinMem}, with the lattice spacings restricted to $a/\lambda < 1/\sqrt{2}$ for subradiant modes to exist.  A similar scheme can also be realized using the bilayer setup, extending the allowed lattice spacing to $a/\lambda < 1$ as we show in Appendix C; yet, it remains inapplicable to the superwavelength regime.

Instead, we show below how the bilayer array system offers more flexibility to this end, by introducing a memory scheme suitable also for superwavelength regimes relevant to tweezer arrays. Importantly, the scheme is based on dynamical switching of the array coupling $\Gamma_q$ to light, via \emph{continuous} tuning of the interlayer spacing $a_z$; namely, it completely relies on new options opened by the non-symmetric operation of the bilayer array and the theory developed above. In addition, the fidelity of the memory is shown to follow the interface efficiency $r_q$ whose optimization was discussed above, again highlighting the universality of this figure of merit for quantum tasks.

In the following, we begin by presenting the general concepts underlying our approach and then discuss their realization.

\subsection{Memory protocol}
Quantum memories rely on efficient coupling $\Gamma_q$ between the target photon mode and a stable atomic degree of freedom $\hat{P}_q$. An additional key ingredient, however, is the ability to switch this coupling on and off, for the storage and retrieval of the converted excitations. This requires the dynamical control over $\Gamma_q(t)$, between large and vanishingly small values, a task which seems natural for the bilayer array interface: Eq. (\ref{effParamarray}) reveals that $\Gamma_q(t)$ is completely determined by the interlayer spacing $a_z$, reaching $\Gamma_q(t)=0$ for the Bragg condition at which $\hat{P}_q$ is a dark mode. So, continuous tuning of the spacing $a_z$ yields the required dynamical control over $\Gamma_q(t)$.

Notably, changing $a_z$ does not change the eigenmodes $\hat{P}_q$ from Eq. (\ref{Pq}) but only their eigenvalues $\Gamma_q(t)/2-i\Delta_q(t)$, with $\gamma_{q,\text{loss}}$ taken independent of $a_z$ considering that diffraction-order losses are eliminated up to residual individual-like contributions $\gamma_s$ (Sec. IV). Therefore, we can choose to couple light to one of the eigenmodes, thus fixing $q$, and control $\Gamma_{q}(t)$ by varying $a_z$ in time. In principle, this should involve the motion of one of the array layers via a continuous uniform shift of its optical traps. We assume that the rate of this shifting is slow enough for the atoms to remain trapped in the same lattice structure, such that $\hat{P}_q$ remains an eigenmode. The conditions that realize this adiabatic operation are discussed in Sec. V B below.

After establishing the principle of an atomic coherence $\hat{P}_q$ whose coupling $\Gamma_q(t)$ varies in time according to Eq. (\ref{effParamarray}) with $a_z(t)$, and whose residual losses $\gamma_{q,\text{loss}}=\gamma_s$ are constant, we proceed to analyze the memory fidelity by studying the storage efficiency. To this end, we consider an input field in the target mode $q$ that contains excitations only in a single temporal (longitudinal) mode $h_{0}(t)$, satisfying the normalization $\intop_{0}^{T}dt\left|h_{0}(t)\right|^{2}=1$ within the protocol duration $T$. Taking, for simplicity, a single-photon excitation, $\langle {\cal E}_{q,\text{in}}^{\dagger}(t'){\cal E}_{q,\text{in}}(t)\rangle =h_{0}^{*}(t')h_{0}(t)$, we solve Eq. (\ref{EOM}) for $\hat{P}_{q}$ and obtain the efficiency of excitation storage,
\begin{eqnarray}
r_{q,s}\equiv \frac{\left\langle \hat{P}_{q}^{\dagger}(T)\hat{P}_{q}(T)\right\rangle }{\intop_{0}^{T}dt\left\langle \hat{{\cal E}}_{q,\text{in}}^{\dagger}\hat{{\cal E}}_{q,\text{in}}\right\rangle }
=r_f\left|\intop_{0}^{T}dt\cdot f(t)h_{0}(t)\right|^{2},
\label{rs}
\end{eqnarray}
with $f(t)=e^{-\intop_{t}^{T}dt'[\Gamma_{q}(t')+\gamma_s+2i\delta-2i\Delta_{q}(t')]/2}\sqrt{\Gamma_{q}(t)/r_f}$ and $r_f$ being the normalization that guarantees $\intop_{0}^{T}dt\left|f(t)\right|^{2}=1$.

Since both the input-field pulse $h_0(t)$ and the function $f(t)$ are normalized with respect to the integral $\int_0^Tdt$, the maximal memory efficiency $r_{q,s}$ is given by $r_f$ and is obtained for $f(t)=h_0^{\ast}(t)$. This can be arranged by noting that the shape of $f(t)$ is determined by the time-dependence of the spacing $a_z(t)$, which can be appropriately chosen to yield $f(t)=h_0^{\ast}(t)$, yielding efficiency $r_{q,s}=r_f$. In turn, $r_f$ is determined by the normalization of $f(t)$ and hence by $a_z(t)$. To analyze the limits on $r_f$, consider e.g. the mode $q=0$, such that we require $a_z(0)=\mathbb{N}\lambda$ and $a_z(T)=a_z(0)+\lambda/2$ to tune the coupling from maximal, at the beginning of the storage process ($t=0$), to zero at its end ($t=T$) [Eq. (\ref{effParamarray})]. For concreteness, consider the case $a_z(t)/\lambda=1+\frac{1}{2} e^{-(T-t)/\tau}$, which approximates the required tuning with a typical ``switch time" $\tau < T$. Assuming the coupling rate is faster than the pulse duration, $\Gamma_{\text{1D}} T \gtrsim 1$, we obtain the efficiency in the generic form (Appendix A),
\begin{eqnarray}
r_{q,s}=r_f \approx 1-\frac{\gamma_s}{2\Gamma_{\text{1D}}}-B \gamma_s \tau.
\label{rf}
\end{eqnarray}
Here, the coefficient $B$ is a weakly-dependent function of $\Gamma_{\text{1D}}$, given by $B\approx (3/4)(\tau\Gamma_{\text{1D}})^{-1/3}$ for the specific $a_z(t)$ from above.

\begin{figure}[t]
  \centering
  \includegraphics[width=\columnwidth]{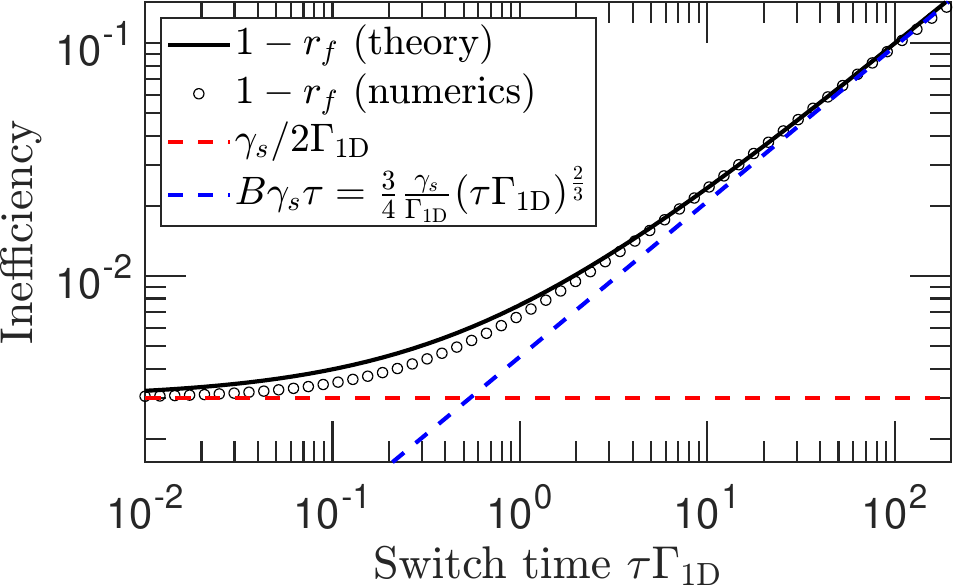}
  \caption{Storage inefficiency $1-r_f$ for the memory protocol with $a_z(t)/\lambda=1-\frac{1}{2} e^{-(T-t)/\tau}$, plotted as a function of the scaled ``switch time" $\tau\Gamma_{1D}$. The theoretical result Eq. (\ref{rf}) (black solid line) exhibits excellent agreement with the numerical estimation from direct integration of the protocol dynamics [Eq. (\ref{g})] (black circles), noting the asymptotic scalings for small and large $\tau\Gamma_{1D}$ corresponding to the second and third terms in Eq. (\ref{rf}) (red and blue dashed lines). Parameters: pulse duration $T\Gamma_{1D}=1000$, optimal loss ratio $\gamma_s/2\Gamma_{1D}=0.003$ (corresponding to residual diffraction-order losses of a tweezer array with $N\sim 4000$ atoms, see Fig.~\ref{Fig3}).}\label{Fig_rf}
\end{figure}

If we ignore the last term in (\ref{rf}), we obtain a memory efficiency that reaches the optimally achievable efficiency $r_q$ of the quantum interface, with maximal coupling $\Gamma_q=2\Gamma_{\text{1D}}$ and minimal losses $\gamma_{q,\text{loss}}=\gamma_s\ll \Gamma_{\text{1D}}$  [Eq. (\ref{rq})]. This situation occurs only if the first loss term in (\ref{rf}) is the dominant one, requiring $(B\tau)^{-1}\gg \Gamma_{\text{1D}}$, i.e. that the switch time is faster than the radiative coupling. In practice, this may violate the adiabatic operation we assumed, as further discussed below. The opposite limit, where the second loss term is dominant, yields the effective loss $\gamma_{q,\text{loss}}/\Gamma_q=B \gamma_s \tau \equiv G \gamma_s/(2\Gamma_{\text{1D}})$, which presents losses amplified by a factor $G=2B \tau\Gamma_{1D}\approx (3/2)(\tau\Gamma_{1D})^{2/3}$ with respect to the optimum case. We plot the resulting inefficiency $1-r_f$ from Eq. (\ref{rf}) in Fig.~\ref{Fig_rf}, using relevant loss parameters deduced from Fig.~\ref{Fig3}. We observe excellent agreement with a numerical estimate obtained by the direct integration of the protocol dynamics (Appendix A), also noting the two asymptotic limits discussed above. These results again highlight how $r_{q}$ fundamentally quantifies the efficiency bounds of quantum operations supported by the array.

\subsection{Realization with moving atoms}
The dependence of the coupling rate $\Gamma_q$ on $a_z$ considered above is readily realized in both the subwavelength and superwavelength bilayer systems discussed in Secs. III and IV, as per Eq. (\ref{effParamarray}). Considering the loss, we assumed individual-like losses independent of $a_z$, $\gamma_{q,\text{loss}}=\gamma_s$; for the subwavelength bilayer array, where all diffraction orders $\mathbf{m}\neq \mathbf{0}$ are non-radiative and $\gamma_{q,\text{diff}}=0$, this is indeed always the case.

In contrast, for the superwavelength bilayer array, a contribution $\gamma_{q,\text{diff}}$ in the losses exists in principle and, crucially, also depends on $a_z$ [Eq. (\ref{G_diff})]. Keeping these losses eliminated while continuously varying $a_z$, will then require continuously varying the intralayer lattice spacing $a$ as well: we recall that configurations $(a_z,a)$ that ensure $\gamma_{q,\text{diff}}=0$ must remain on the resonant curves, Eq. (\ref{curves}), requiring to change $a$ in tandem with any change in $a_z$.

The required temporal changes in array spacings entail the motion of atoms by moving their traps --- either globally for an entire layer as in $a_z$, or locally within a layer to change the lattice constant $a$. In order to keep the atoms trapped in the desired positions and avoid effective heating during these changes, the rates $|\dot{a}_z/a_z|,|\dot{a}/a|\sim \tau^{-1}$ should be smaller than the trap frequency of the laser-trapping potential. In current platforms, this limits the rate to about $\tau^{-1}\sim 10$ kHz, as was recently employed in a reconfigurable tweezer-array processor \cite{LUKbluv}. Considering superwavelength lattice spacings $a/\lambda>1.5$ such that $\Gamma_{\text{1D}}/\gamma \sim (3/4\pi)(a/\lambda)^2<1/10$, we then get for e.g. rubidium atoms $\tau\Gamma_{\text{1D}}\sim 100$.
With the parameters of Fig.~\ref{Fig_rf}, this yields storage efficiencies $r_f$ exceeding 0.9. Notably, the achievable atom number $N$ and motion rate $\tau^{-1}$ are both likely to increase in the coming years, considering their importance for tweezer-array quantum technologies. In turn, this makes our memory scheme more viable, as the inefficiency $1-r_f$ decreases with both $N$ and $\tau^{-1}$: the former via the scaling $\gamma_s\sim N^{-1}$ of residual diffraction-order losses (Fig. 3), and the latter via the scaling $\tau ^{2/3}$ of the last term in Eq. (\ref{rf}) (Fig. 7).

For superwavelength arrays, the need for temporal changes also in the intralayer lattice spacing $a$ (performed in tandem with $a_z$) formally poses an additional adiabatic constraint, found in Appendix B as $|\dot{a}/a|\sim \tau^{-1}\ll \Gamma_{\text{1D}}$. Nevertheless, considering the parameters discussed above, with $\tau\Gamma_{\text{1D}}\sim 100$, this constraints is already naturally satisfied.

We note that the analysis above concerns only the storage and retrieval processes themselves; additional limitations may arise during storage time, when the excitations reside in the collective dark state. The dark state lifetime is limited by residual non-collective decoherence and by finite-size or disorder effects \cite{Uni}. Additionally, momentum-dependent dephasing can occur during storage \cite{YelinMem}, when different momentum components of the stored excitation acquire relative phases during storage time. This latter effect can be diminished by considering spatial confinement of the dipolar excitation wavepacket on the array via light shifts \cite{WildCav}.

\section{Discussion}
We presented a general framework for non-symmetric quantum interfaces, showing that the efficiency of quantum tasks can be directly determined from classical scattering observables --- the reflection and transmission. We demonstrated a concrete realization in bilayer atomic arrays with arbitrary interlayer spacing, going beyond symmetric Bragg configurations. Exploiting the non-symmetric regime, we showed enhanced interface efficiencies in tweezer arrays by cancellation of both first- and second-order diffraction losses, and explored new quantum-memory schemes based on collective dark states.

Going forward, the developed framework should serve as the basis for a systematic exploration of non-symmetric quantum interfaces and their potential benefits. In this respect, two related aspects to pursue are: (i) the extension of the formalism to array geometries beyond the bilayer case considered here; and (ii) the study of specific quantum tasks and applications that rely on nonlinearities of non-symmetric interfaces, e.g. for the generation of entanglement and correlated quantum states of light.

\begin{acknowledgments}
We acknowledge financial support from the Israel Science Foundation (ISF), the Minerva Stiftung with funding from the Federal German Ministry for Education and Research, the Israel Council for Higher Education (VATAT), the US-Israel Binational Science Foundation (BSF) and US National Science Foundation (NSF), the Center for New Scientists at the Weizmann Institute of Science, the Helmsley Charitable Trust, and the Estate of Louise Yasgour. This research is made possible in part by the historic generosity of the Harold Perlman Family.
\end{acknowledgments}


\appendix
\section{Analysis of memory efficiency}
Here we elaborate on the result (\ref{rf}) of the memory efficiency $r_{s,q}=r_f$. Recalling the definition of $f(t)$ below Eq. (\ref{rs}), $r_f$ is fixed by the normalization condition of $f(t)$ and is therefore given by
\begin{eqnarray}
r_f&=&\intop_{0}^{T}dte^{-\intop_{t}^{T}\left[\Gamma_{q}\left(t'\right)+\gamma_{s}\right]dt'}\Gamma_{q}\left(t\right)
\label{g}
\end{eqnarray}
For $a_z(t)/\lambda=1-\frac{1}{2} e^{-(T-t)/\tau}$ discussed in the main text, we express the integrand in terms of cosine-integral functions and evaluate the integral for $r_f$ numerically in representative cases and limits, finding agreement with the expression in Eq. (\ref{rf}) with the slowly varying coefficient $B\approx (3/4)(\Gamma_{\text{1D}}\tau)^{1/3}$.


To make sense of this result, we considered a simplified model for tuning $a_z$ which follows the same generic timescales: $a_z$ remains at $\lambda$ for the interval $t\in[0,T-\tau]$ and then immediately switches to $\lambda/2$  for the remaining interval  $t\in[T-\tau,T]$. With this model for $a_z$ we obtained analytically the result (\ref{rf}), however, this time with $B=1$, suggesting the following intuitive picture: as long as the coupling is on, the inefficiency $1-r_f$ is that due to the balance of coupling and loss, as expressed by the first loss term in (\ref{rf}), $\gamma_s/(2\Gamma_{\text{1D}})$. In turn, for the duration $\tau$ at which the coupling is going to zero, there remains only the loss $\gamma_s$, contributing the second loss term $\sim\gamma_s\tau$ to $r_f$.

This result is therefore quite generic, with the specifics of the temporal switching from $\lambda$ to $\lambda/2$ entering into the details of the weak dependence of the coefficient $B$ on $\Gamma_{\text{1D}}$: ranging from no dependence at all in the case of abrupt switching at $t=T-\tau$, to $(\tau\Gamma_{\text{1D}})^{-1/3}$ for the gradual exponential switch considered in the main text. For gradual switching, the weak scaling $B\sim (\Gamma_{\text{1D}}\tau)^{-1/3}$ appears to be rather generic as well: this is seen by considering the limiting case of linear tuning, $a_z(t)/\lambda=1+t/(2T)$, where we again obtain $B\approx (3/4) (\Gamma_{\text{1D}}\tau)^{-1/3}$.

\section{Additional adiabatic constraint for superwavlength arrays}
In Sec. V B we discuss the realization of the memory protocol via temporal changes in the interlayer spacing $a_z$, recalling that for supwewavlength array, this requires changing in tandem also the intralayer lattice spacing $a$. Formally, this adds another condition for the system to evolve adiabatically, beyond that related to the finite trap potential.
In particular, while the eigenmodes $\hat{P}_q$ of the bilayer array remain unchanged in terms of the collective dipoles $\hat{P}_{l}$ of the constituent layers, the individual-layer dipoles $\hat{P}_{l}$ themselves become time-dependent via their dependence on $a(t)$. For large arrays, the planar eigenmodes of each layer are approximate transverse momentum modes, $\hat{P}_{l,\boldsymbol{k}_{\perp}}=\frac{1}{\sqrt{N}}\sum_{\mathbf{n}}|g\rangle \langle e|_{\mathbf{n}l}e^{-i\boldsymbol{k}_{\perp}\boldsymbol{r}_{\boldsymbol{n}l}}$, with $\boldsymbol{k}_{\perp}=(k_x,k_y)$ spanning the Brillouin zone of the 2D lattice. While the infinite-layer theory assumed plane waves and hence $\hat{P}_l=\hat{P}_{l,\boldsymbol{k}_{\perp}=0}$, practical target fields have a finite beam waist, and hence a finite transverse-momentum spread around $\mathbf{k}_{\perp}=0$, which can mix due to the finite change rate $\dot{a}/a$. Considering the typical radiative widths of these modes, given by $\sim \Gamma_{\text{1D}}$, the adiabatic condition to prevent this mixing is then $|\dot{a}/a|\sim \tau^{-1}\ll \Gamma_{\text{1D}}$. In practice, however, this condition does not result in any further limitations beyond those related to the trapping potential, as discussed in the main text.

\section{Realization of the memory protocol via light shifts}
For subwavelength arrays, a quantum-memory protocol, similar to that from Sec. V, can be realized also without changing the interlayer spacing $a_z$, by introducing layer-dependent light shifts instead. Specifically, we consider applying uniform light shifts on each layer $l=1,2$, such that the corresponding collective dipoles $\hat{P}_{1,2}$ acquire energy shifts of opposite signs, $\Delta_{\text{LS}}=\Delta_{1}=-\Delta_{2}$. The opposite light shifts on the two layers break the symmetry between layers, thus inducing a coherent coupling between the collective modes $\hat{P}_{0}$ and $\hat{P}_{\pi}$ defined in Eq. (\ref{Pq}). In the Bragg-symmetric configuration, one mode becomes fully bright while the other becomes fully dark, thus serving as a basis for the memory protocol. In particular, denoting the dark mode with $\Gamma_{0}=0$ by $\hat{S}\equiv\hat{P}_{0}$ and the bright mode with $\Gamma_{\pi}=2\Gamma_{\text{1D}}$ by $\hat{P}\equiv\hat{P}_{\pi}$ , the equations of motion are recast as
\begin{eqnarray}
\dot{\hat{P}}
&=& \left[-\frac{2\Gamma_{\text{1D}}+\gamma_{s}}{2}
  + i(\delta-\Delta_{\text{bright}})\right]\hat{P}
  + i\Delta_{\text{LS}}(t)\hat{S} \nonumber \\
& &+ i\sqrt{2\Gamma_{\text{1D}}}\hat{\mathcal E}_{\text{in}}(0,t)
  + \hat{F}_{\text{bright}}(t), \nonumber  \\[6pt]
\dot{\hat{S}}
&=& \left[-\frac{\gamma_{s}}{2}
  + i(\delta-\Delta_{\text{dark}})\right]\hat{S}
  + i\Delta_{\text{LS}}(t)\hat{P} + \hat{F}_{\text{dark}}(t). \nonumber \\
\label{EOM_LS}
\end{eqnarray}
These resulting equations of motion are directly equivalent to those of a three-level-atom quantum memory analyzed in Ref.~\cite{Uni,ref59}, with the light shift assuming the role of the control field $\Omega$ from Fig. 4. Adiabatic elimination of the bright mode $\hat{P}$, yields an effective equation of motion for $\hat{S}$ of the same form as Eq. (\ref{EOM}), with an effective, time-dependent coupling to the input field whose time dependence is controlled by the light shift. Following a similar analysis to that of Sec.~V A, the storage fidelity is found to be the same as Eq.~(\ref{rf}), with the only modification being that the first loss term is now multiplied by two, hence given by $\gamma_{s}/\Gamma_{\text{1D}}$. Importantly, this memory scheme is limited to operation in the Bragg-symmetric configuration and is therefore restricted to subwavelength arrays, since in the superwavelength regime the diffraction-order losses cannot be simultaneously eliminated for both the bright and dark modes, thereby limiting the achievable efficiency.



\begin{thebibliography}{10}
\bibitem{note}
{For the Bragg solutions, we see from Eq. (12) that one mode is always dark,
  with coupling to light $\Gamma_q=0$, while the other always maximally bright,
  with $\Gamma_q=2\Gamma_{\text{1D}}$. Therefore, the dark mode is always
  irrelevant for light interfacing even though it has vanishing losses,
  $\gamma_{q,\text{loss}}=0$ (being on a resonant curve).}
\end{thebibliography}

\end{document}